\documentclass[aps,english,prl,twocolumn,showpacs,superscriptaddress]{revtex4-1}
\usepackage{graphicx}
\usepackage{natbib}
\usepackage{amssymb}
\usepackage{babel}
\usepackage{epstopdf}
\usepackage{subfig}
\usepackage{color}
\usepackage{caption}
\begin{document}

\bibliographystyle{apsrev}

\title{Evidence for a Finite-Temperature Insulator}

\author{M. Ovadia}
\affiliation{Department of Condensed Matter Physics, The Weizmann Institute of Science, Rehovot 76100, Israel.}
\affiliation{Present Address: Department of Physics, Harvard University, Cambridge, Massachusetts 02138, United States.}
\author{D. Kalok}
\affiliation{Department of Condensed Matter Physics, The Weizmann Institute of Science, Rehovot 76100, Israel.}
\author{I. Tamir}
\affiliation{Department of Condensed Matter Physics, The Weizmann Institute of Science, Rehovot 76100, Israel.}
\author{S. Mitra}
\affiliation{Department of Condensed Matter Physics, The Weizmann Institute of Science, Rehovot 76100, Israel.}
\author{B. Sac\'ep\'e}
\affiliation{Department of Condensed Matter Physics, The Weizmann Institute of Science, Rehovot 76100, Israel.}
\affiliation{Univ. Grenoble Alpes, Institut NEEL, F-38042 Grenoble, France.}
\affiliation{CNRS, Institut NEEL, F-38042 Grenoble, France.}
\author{D. Shahar}
\email{dan.shahar@weizmann.ac.il.; Correspond author}
\affiliation{Department of Condensed Matter Physics, The Weizmann Institute of Science, Rehovot 76100, Israel.}

\begin{abstract}
In superconductors the zero-resistance current-flow is protected from dissipation at finite temperatures ($T$) by virtue of the short-circuit condition maintained by the electrons that remain in the condensed state. 
The recently suggested finite-$T$ insulator and the \textquotedblleft superinsulating" phase are different because any residual mechanism of conduction will eventually become dominant as the finite-$T$ insulator sets-in. If the residual conduction is small it may be possible to observe the transition to these intriguing states. 
We show that the conductivity of the high magnetic-field insulator terminating superconductivity in amorphous indium-oxide exhibits an abrupt drop, and seem to approach a zero conductance at $T<0.04$ K. We discuss our results in the light of theories that lead to a finite-$T$ insulator. 
\end{abstract}

\maketitle

In 2005, two theoretical groups\cite{basko,GornyiPrl} considered a disordered, strongly interacting, many-body system of electrons that is not coupled to an external environment (phonons). They posed the fundamental question of whether thermal excitations, which are essential to the mechanism of charge transport, can equilibrate via the interaction with the electron bath or stay frozen as a consequence of, what they termed, the many-body localization (MBL). Their analyses indicated that in such a system an insulating, zero conductance ($\sigma$), state is identified at finite-$T$ up to a well-defined critical $T$, $T^{*}$. Numerical calculations\cite{OganesyanPrb,huseprb87} based on the analytical approach of Ref.\cite{basko} provide ambiguous results regarding the existence of such a phase at nonzero $T$'s. 

In order to experimentally search for this finite-$T$ insulator, it was later suggested\cite{baskoprb}, one should look in disordered systems in which the electrons decouple, at low $T$, from the phonons. A clear signature of this decoupling is the appearance of discontinuities in the current-voltage ($I$-$V$) characteristics\cite{sanprb53} that result from bi-stability of the electrons $T$ ($T_e$) under $V$-bias conditions. 

We focus on highly disordered superconductors that, at high magnetic-field ($B$), undergo a superconductor-insulator transition (SIT)\cite{goldmanpt51,physupekhi}.
The SIT is a quantum phase transition\cite{sondhirmp} that can be driven by $B$\cite{HebardPrl,kapitulnikprl74,BaturinaJETP}, disorder\cite{Shaharprb}, thickness\cite{haviprl62}, gate voltage\cite{goldmanprl94} or other parameters in the Hamiltonian. It is observed in variety of systems\cite{HebardPrl,kapitulnikprl74,BaturinaJETP,haviprl62,bounatmat11} and by various experimental techniques\cite{HebardPrl,benjaminprl101,craneprb752}.
 
In the $B$-driven SIT the superconductor goes into an insulating phase at a critical $B$, $B_{C}$. In many cases\cite{paalanenprl69,GantmakherJETP,murthyprl2004,vallesprl103,benjaminprl101,kapitulnikprl74}
strong insulating behavior is seen only over a narrow range of $B$ to form an \textquotedblleft insulating peak" (see figure 1). Both theoretical\cite{FeigAnnals,YonatanNat} and experimental\cite{GantmakherJETP,murthyprl,vallesprl103} studies associate the insulating peak with Cooper-pair localization.

\begin{figure}
\includegraphics  [width=8.5 cm] {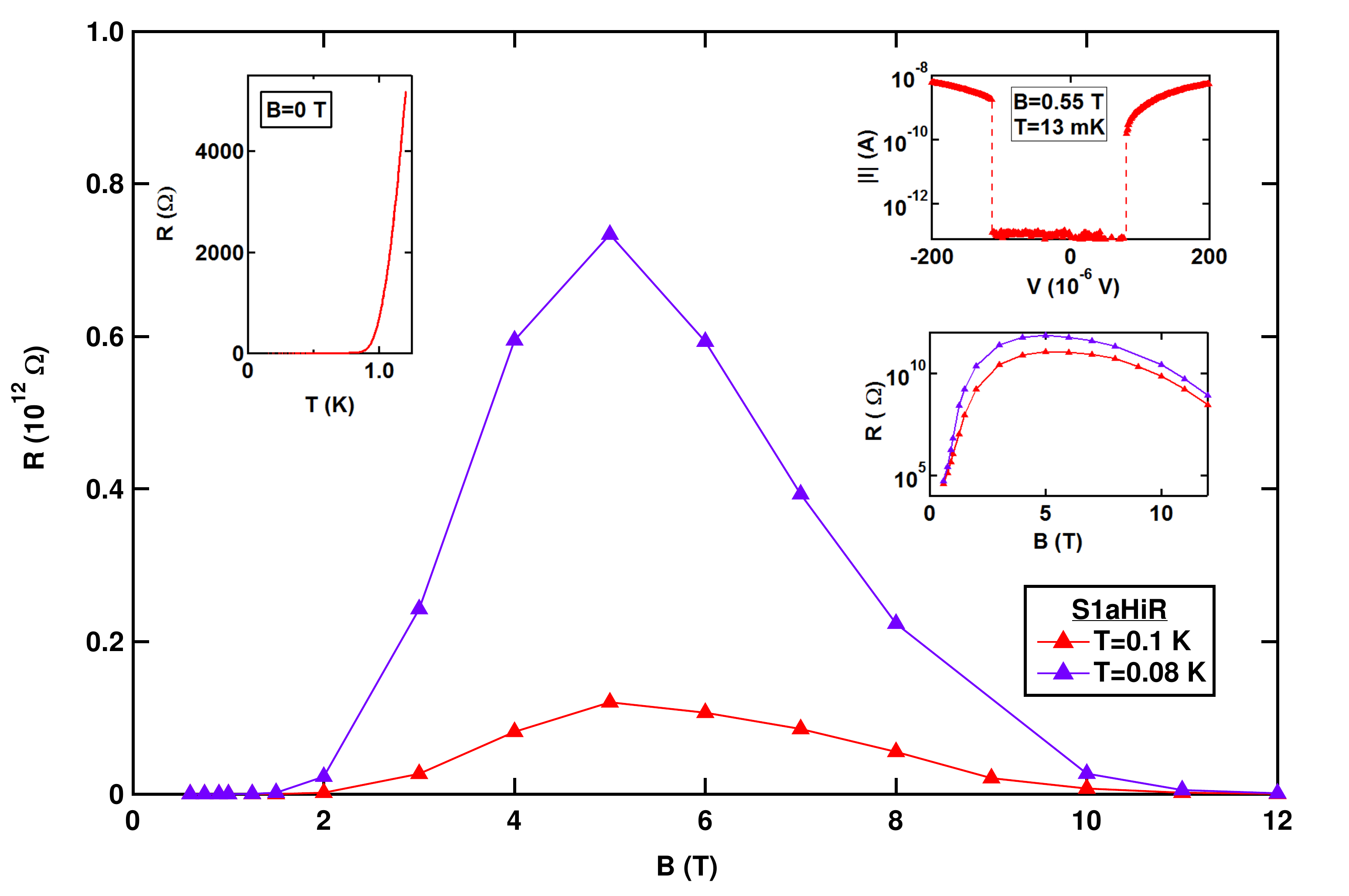}
\caption{{\bf Insulating peak.} $R(B)$ isotherms from $B$=0.5 to 12 T, measured at $T$=0.1 K (red) and $T=0.08$ K (blue). Both show the peak at $B$=5 T. The data (triangles) were extrapolated from $I$-$V$ scans. Data is taken from our main results shown in figure 2. [Left inset]: Superconducting phase transition at $B=0$ with $T_c\approx1.1$ K.  [Top right inset]: $I$-$V$ characteristic measured at $B$=0.55 T and $T$=13 mK showing the abrupt jump of more than 4 orders of magnitude in $I$ at a particular threshold $V$. [Bottom right inset]: The same set of data as in the main figure using log scale for $R$. In all figures, except for the left inset, the lines are guides to the eye.} 
\label{peak}
\end{figure}

To characterize this $B$-induced insulating peak, we\cite{murthyprl} studied its $I$-$V$ characteristics 
and found that they exhibit a discontinuous jump in $I$ of more than 4 orders of magnitude as a threshold $V$, $V_{th}$, is exceeded (see top right inset of figure 1). This finding\cite{BaturinaPRL,KalokJP} was theoretically linked\cite{vinokurnat} to the formation of a \textquoteleft superinsulating' state that in a manner akin, but opposite, to superconductivity is characterized by an abrupt vanishing of $\sigma$ at low $V$-bias. 

An alternative view of the discontinuous $I$-$V$ characteristics was offered by Altshuler \textit{et al.}\cite{borisprl} who analyzed the steady state heat balance in the insulating-peak region under $V$-bias. They suggested that the $I$ jumps resulted from bi-stability of $T_{e}$ that, at low $T$, can be very different from the $T$ of the host phonons ($T_{ph}$). We followed this theoretical work with a systematic study and obtained a good agreement\cite{maozprl}. 
We were also able to estimate the $T$ dependence of the $e$-$ph$ scattering rate, $\tau_{e-ph}$, on the high $B$ side of the insulating peak and found a rather strong dependence of $\tau_{e-ph}\sim T^{-4}$, which is in agreement with the modified dirty metal model\cite{borisprl,kravtsovprl111}. 
The success of this theoretical description provides an essential indication that, in our regime of measurements, the electrons are decoupled from the phonons.

The realization that our samples exhibit a strongly $T$-dependent insulating behavior with diminishing $e$-$ph$ coupling motivated us to conduct a systematic study of their Ohmic transport at very low $T$ ($T<0.3$ K). In order to achieve that, we had to greatly improve our ability to measure very high sheet resistance ($R$). While our earlier studies\cite{murthyprl2004} were limited to $R$ up to $10^9$ $\Omega$, several improvements (described in the supplementary materials) extended the range of our measurements to $10^{12}$ $\Omega$. 
These improvements enabled the results that follow.

The data presented here are obtained from the sample S1aHiR, a thin film of a:InO, patterned in Hall bar geometry, $0.5\times 0.25$ mm$^2$ in size. The sample is superconducting at $B$=0 with a $T_c\approx1.1$ K (see left inset of figure 1) and undergoes a $B$-driven SIT. In figure 1 we show two isotherms of $R$ in the insulating region, as a function of $B$ from 0.5 to 12 T, at $T$=0.08 and 0.1 K. Both show the insulating peak at 5 T. Due to technical reasons we were unable to pinpoint the $B_c$ of our sample but located it to be between 0.16 and 0.4 T. The sample exhibited the thermal bi-stability in the insulating phase as evident by a typical 
$I$-$V$ characteristic\cite{maozprl}, at $B$=0.55 T and $T$=13 mK, shown in top right inset of figure 1.

Our main results are presented in figure 2 where we plot the $T$-dependence of $R$ at various $B$'s, from 0.5-12 T, spanning the insulating peak. Depending on the $R$-range, measurements were done using two different techniques. For the moderate-$R$ range ($R<10^{8}$ $\Omega$) data were obtained by continuous two-terminal measurements (solid lines), whereas for $R>10^{8}$ $\Omega$ each datum (marker) was obtained from a full $I$-$V$ scan (see methods). The dashed lines joining the markers are guides to the eye. 

\begin{figure}
\includegraphics [width=8.5 cm] {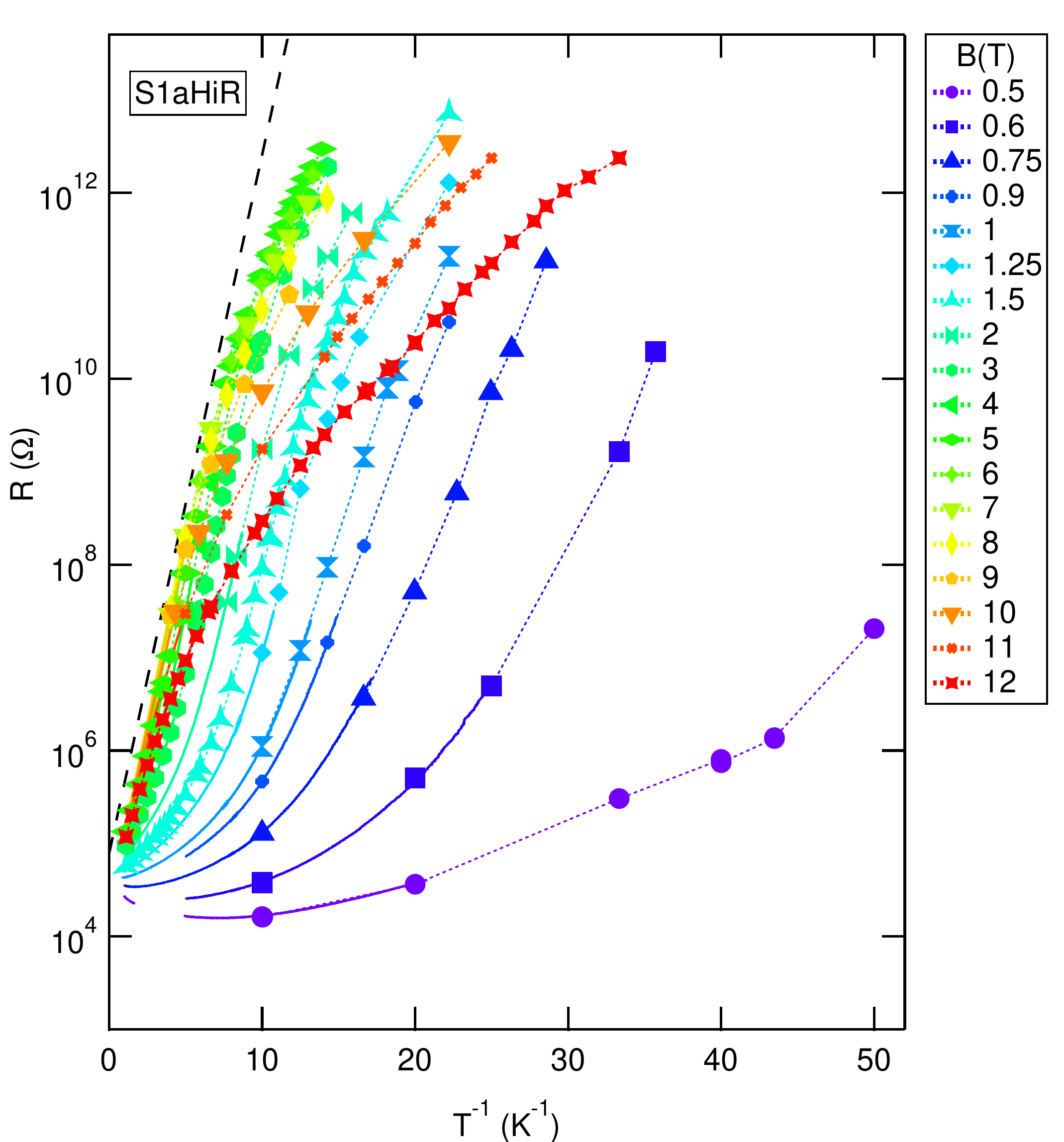}
\caption{{\bf \boldmath$T$ dependence of \boldmath$R$ using Arrhenius mapping.} $R$ (in log scale) vs. $1/T$ at different $B$'s ranging from 0.5 to 12 T. The solid lines represent data acquired by two-terminal measurements, while data obtained from $I$-$V$ scans are shown as markers. The dashed lines joining the markers are guides to the eye. The dashed black line demonstrates how an activated behavior should appear in an Arrhenius plot.} 
\label{MainRT}
\end{figure}

Based on earlier studies which were limited to a much lower $R$-range, we were anticipating activated behavior\cite{murthyprl2004,BaturinaPRL} and adopted an Arrhenius form to present our data. However, the broad range of $R$ in this study brings about the observation of clear deviations from activated transport. While the low $R$ ($R<10^6$ $\Omega$) data are still consistent with activated behavior (for reference we added a dashed black straight line, indicating activated behavior in figure 2) the high $R$ data, offering several orders of magnitude broader range, clearly are not. 

The deviations, seen in all $B$ values of figure 2, crucially differ depending on the value of $B$. At the high $B$'s, the convex shape of the curve indicates sub-activation behavior. This behavior is illustrated in figure 3(a) where $R (B$=12 T$)$ is plotted (in red), using a logarithmic scale, vs. $T^{-1/2}$. The data convincingly follow a straight line over our full $T$-range indicating,
\begin{equation}
R(T)=R_{ES}exp[(\frac{T_{ES}}{T})^{1/2}].
\label{e1}
\end{equation}
This is consistent with the Efros-Shklovskii (ES) variable range hopping (VRH) mechanism of transport\cite{efros}.  $T_{ES}$ and $R_{ES}$ are the ES temperature ($T_{ES}$=14.8 K) and pre-factor respectively. This dependence holds, with increasing $T_{ES}$, for $B$'s down to the peak position (at $B$=5 T, $T_{ES}$= 23.6 K).

\begin{figure} [p]
\includegraphics [width=12 cm] {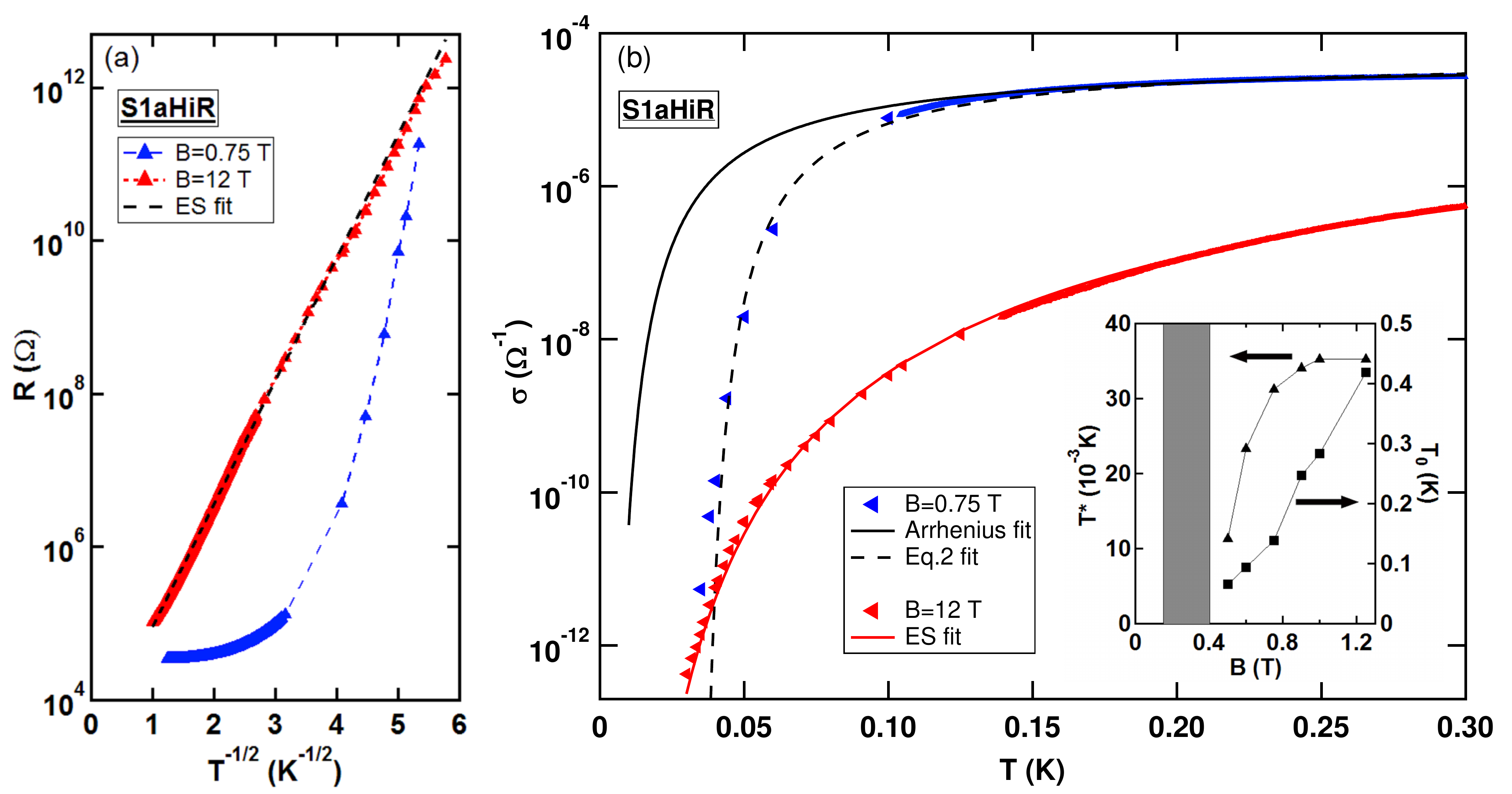}
\caption{{\bf Mapping \boldmath$T$ dependence of \boldmath$R$ and $\sigma$.} (a) ES type mapping of $T$ dependence of $R$. $R$ (in log scale) as a function of $T^{-1/2}$:  At $B=12$ T (in red), including ES fit (in dashed black line) and at $B=0.75$ T (in blue). The low $B$ data clearly deviates from the ES type, indicating that at lower $B$s electronic transport in our system do not follow the ES VRH.
(b) Vanishing conductivity at non-zero $T$. The variation of $\sigma$ (in log scale) as a function of $T$ at $B$=0.75 T. The solid black line is a fit using Arrhenius form. The dashed black curve is the fit to Eq.\ref{e2}. For reference we add $\sigma(T)$ taken at $B$=12 T (red triangles) where ES dependence holds (red curve). In both (a) and (b) the solid lines represent data acquired by two-terminal measurements, while data obtained from $I$-$V$ scans are shown as triangles connected by dashed lines as guide for the eye. [Inset] The variation of $T^{*}$ (left axis) and $T_{0}$ (right axis) [see Eq.\ref{e2}] with $B$. The values were obtained by fitting of our experimental data described in figure  2 with Eq.\ref{e2}. The shaded region indicates the $B$ values $B_{c}$ can take.} 
\label{sigT}
\end{figure}

The picture changes dramatically at lower $B$'s, approaching the SIT ($0.5<B<2$ T). An attempt, shown in blue in figure 3(a), to plot data taken in this $B$ range using the ES form clearly fails. A simple activated form is also inadequate as the data clearly appear concave (see figure 2).

The concave curvature evident in the $B<2$ T data of figure 2 signals an unusual, faster than exponential\cite{baturinajetp88}, $R(T)$ dependence. The anomaly is clearly seen when we plot, in figure 3(b), $\sigma$ ($\sigma=\frac{1}{R}$) as a function of $T$ at $B$=0.75 T.
Focusing on the $T<$ 0.3 K range we see that $\sigma$ decreases moderately upon cooling until $T$=0.1 K and then
undergoes a precipitous drop of 6 orders of magnitude to the noise level in our measurement ($\sigma=10^{-12}$ $\Omega^{-1}$). As we stated earlier, our attempts, indicated by the black curve in figure 3(b), to fit these data with an Arrhenius form, failed. For reference we add $\sigma(T)$ taken at $B$=12 T where ES dependence holds (shown in red in that figure).

Our inability to fit the data using an exponential or stretched exponential dependence along with the $e$-$ph$ decoupling we observe in our samples point in the direction of a finite-$T$ insulator\cite{baskoprb}. To test this possibility we fit our data with the following phenomenological form:
\begin{equation}
\sigma(T)=\sigma_{0}exp[- \frac{T_{0}}{T-T^{*}}],
\label{e2}
\end{equation}
which describes the vanishing of the conductivity at finite $T= T^{*}$. The result of our fit is plotted using the black dashed line in figure 3(b), from which we obtain $T_{0}$=0.138 K and $T^{*}$=0.031 K. The data follow this functional form down to $T=0.042$ K and $\sigma=1.3 \times 10^{-10}$ $\Omega^{-1}$, where deviation larger than our measurement accuracy develop. 

In any real system $\sigma$=0 is not a realistic expectation. This is because when $\sigma$ becomes very small other, parallel, channels will carry the electronic current and contribute to $\sigma$. Each such channel will lead to the measured $\sigma$ being higher, and can account for the deviations we observe at $\sigma<1.3 \times 10^{-10}$ $\Omega^{-1}$. These can be due to physical processes within the sample or, possibly, due to leakage currents elsewhere in the measurement circuit. More recently, a theoretical paper utilizing a mean field description to a system near the MBL transition\cite{RahulArx} suggested such deviations should be expected. 

By using Eq.\ref{e2} we do not intend to adhere to a specific theoretical model\cite{GornyiPrl}. It is merely a phenomenological description intended to highlight the unusual aspect of our data: $\sigma(T)$ exhibits a dramatic drop at $T<$0.1 K and appear to approach $\sigma$=0 at a finite $T=T^{*}$. The $B$-dependence of $T^*$ and $T_{0}$ obtained by fitting our data using Eq.\ref{e2} are plotted as the inset in figure 3(b). The shaded region indicates the approximate location of the SIT in this sample. It is worth noting that both $T^*$ and $T_{0}$ seem to approach zero in this region.  

Another way to illustrate the abrupt nature of the conductivity drop near $T^{*}$ is to compare it to the superconductivity transition in one of our disordered a:InO films. In figure 4 we plot $\sigma$ vs. $T$ at $B$=0.75 T for this sample, whereas in the inset we plot $R$ vs. $T$ for sample MInOLa4 at $B$=0 T. Despite the different $T$-range their appearance is remarkably similar: both quantities exhibit a sharp drop over a rather narrow $T$-range. 

\begin{figure} [H!]
\includegraphics [width=8.5 cm] {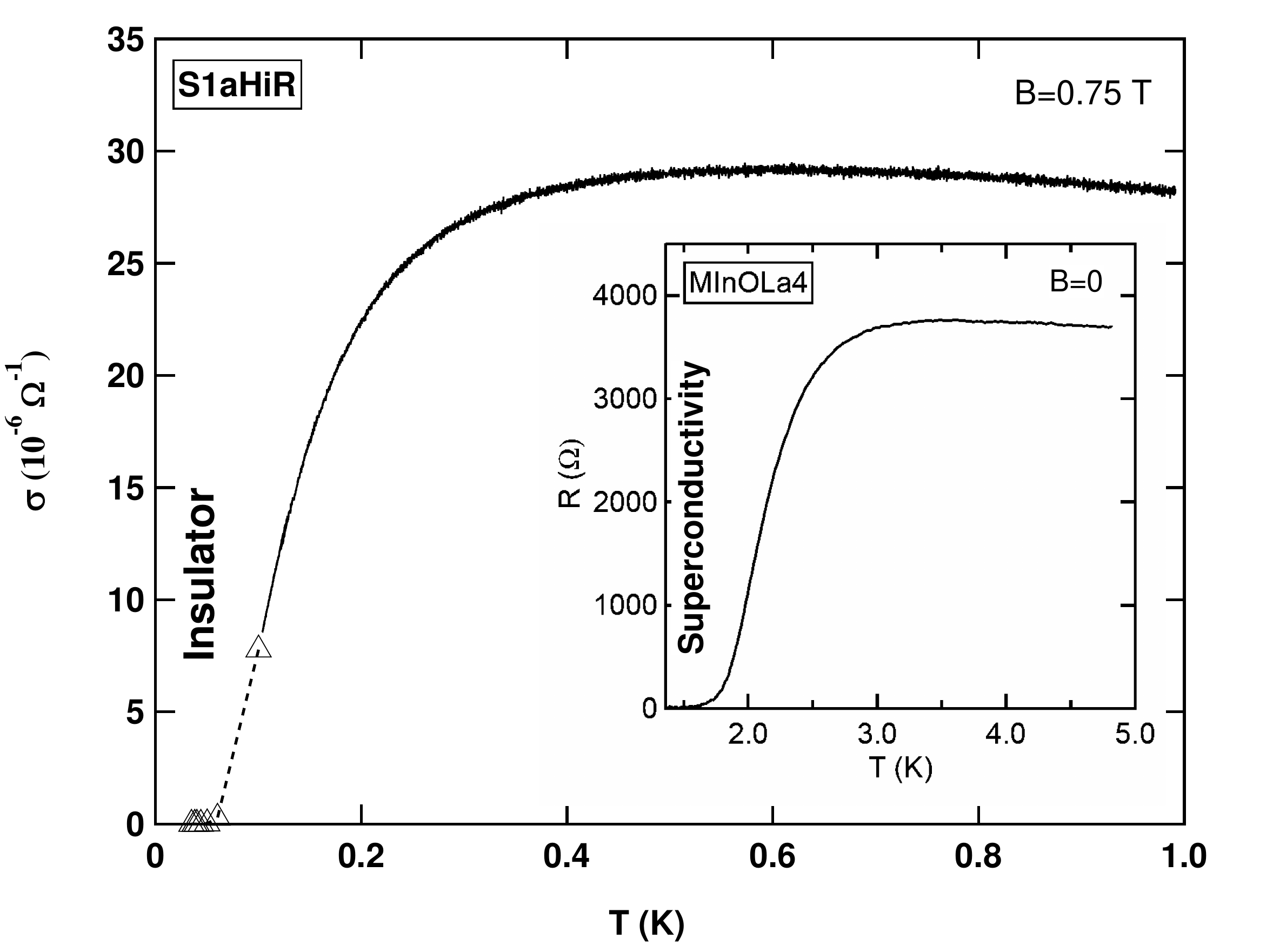}
\caption{{\bf Comparison of Superconductor and finite-\boldmath$T$ insulator.} $T$ variation of $\sigma$ at $B$=0.75 T for this (S1aHiR) sample. The solid lines represent data acquired by two-terminal measurements, while data obtained from $I$-$V$ scans are shown as triangles. [Inset] $R$ at $B$=0 for sample MInOLa4.}
\label{CompareSCandSI}
\end{figure}

It is important to discuss one alternative to Eq. \ref{e2} that, on first sight, appears to agree with our results. At least some of the lower $B$ data of figure 2 can be described, at $T<0.05$ K, by an Arrhenius form indicating activated transport, which results from a mobility gap in the spectrum. A quantitative analysis clearly renders this view inadequate for the following reason. Fitting the $B=0.75$ T data using an Arrhenius form leads to an activation $T$ of 0.91 K. If a mobility gap of such magnitude existed in our system we would expect a much sharper increase in $R$ at $0.91>T>0.05$ K, as seen in the fit presented in the supplementary material. This drop is clearly missing in our data rendering an activated interpretation highly unlikely unless the 0.91 K gap only opens at $T<0.1$ K. We are not aware of a theoretical work predicting such a possibility.

While the new results presented here appear to be in contradiction with earlier findings \cite{murthyprl2004,BaturinaPRL} of activated transport in the peak region, this is not the case: the activation behavior is seen at $T$'s higher than 0.2 K, below which deviations from activation are seen (see figure 2). For these higher $T$'s, where activation is seen, the maximum value of the activation energy is close to $T_C(B=0)$, confirming earlier observations.

The data we are showing here is consistent with transition into a finite-$T$ insulating state. It is tempting to associate this state with the MBL state suggested theoretically\cite{basko, GornyiPrl,OganesyanPrb,huseprb87}. Some of the ingredients are certainly present: our system is highly disordered, strongly interacting and, at the relevant $T$, the electrons decouple from the phonons.

There are other tests that are needed to fully establish the link between our observations and the MBL state chief among which is showing that our electrons are ineffective in reaching equilibrium \cite{GornyiPrl,basko}. This is usually indicated by the presence of long relaxation times in transport. So far, in our experiments, we have not seen such effects but Ovadyahu's group, who study similar materials in a different regime, reported such slow relaxation phenomena\cite{ov1,ov2}.

On the other hand, we recall that the systems in which we observe the transition to the finite-$T$ insulating state are  superconductors at low $B$ and only becomes insulating as $B$ is increased beyond the SIT. Furthermore Cooper-pairing is still dominant in transport even within the insulating regime. While the possible role of Cooper-pairs in forming the finite-$T$ insulator was not considered within the framework of the MBL theories, it was explicitly considered by Vinokur\cite{vinokurnat}  \textit{et al.}, in accordance with the suggested duality\cite{MaozNat} nature of the \textquoteleft superinsulating' state and, more recently, by Feigel'man \textit{et al.}\cite{FeigPrb} who considered the fractal nature of the electronic wave function near a mobility edge and suggested that, if an attractive interaction near the SIT is considered, a finite-$T$ insulator become feasible. More detailed experiments are needed to test the relevance of these theories.

In summary, we have been able to observe an abrupt drop in $\sigma$ by several orders of magnitude occurring at $T<$0.1 K in a:InO thin film near $B$ induced SIT. This has been found to occur at $T$ and $B$ where the electrons decouple from the host lattice phonons.
The measured data cannot be explained using ES model but fit well with the finite-$T$ electron localization down to a certain conductivity.

\section*{Acknowledgments}
We are grateful to  B. Altshuler, I. Aleiner, E. Altman, D. M. Basko, M. Feigel’man, V. Kravtsov, M. M\"{u}ller, Y. Oreg, Z. Ovadyahu, S. Sondhi and V. M. Vinokur for fruitful discussions. This work was supported by the Israeli Science Foundation and the Minerva Foundation with funding from the Federal German Ministry for Education and Research.

\section*{Author contributions}
M.O., D.K. and B.S. prepared the samples and carried out the experiment. M.O., D.K, B.S. and D.S. analyzed the data.  B.S. and D.S. initiated this work. B.S., I.D., S.M. and D.S. wrote the paper. All the authors discussed the results and commented on the manuscript.

\section*{Competing financial interests}
The authors declare no competing financial interests.

\bibliography{s1ahir}

\begin{thebibliography}{38}
\expandafter\ifx\csname natexlab\endcsname\relax\def\natexlab#1{#1}\fi
\expandafter\ifx\csname bibnamefont\endcsname\relax
  \def\bibnamefont#1{#1}\fi
\expandafter\ifx\csname bibfnamefont\endcsname\relax
  \def\bibfnamefont#1{#1}\fi
\expandafter\ifx\csname citenamefont\endcsname\relax
  \def\citenamefont#1{#1}\fi
\expandafter\ifx\csname url\endcsname\relax
  \def\url#1{\texttt{#1}}\fi
\expandafter\ifx\csname urlprefix\endcsname\relax\def\urlprefix{URL }\fi
\providecommand{\bibinfo}[2]{#2}
\providecommand{\eprint}[2][]{\url{#2}}

\bibitem[{\citenamefont{Basko et~al.}(2006)\citenamefont{Basko, Aleiner, and
  Altshuler}}]{basko}
\bibinfo{author}{\bibfnamefont{D.~M.} \bibnamefont{Basko}},
  \bibinfo{author}{\bibfnamefont{I.~L.} \bibnamefont{Aleiner}},
  \bibnamefont{and} \bibinfo{author}{\bibfnamefont{B.~L.}
  \bibnamefont{Altshuler}}, \bibinfo{journal}{Annals of Phys.}
  \textbf{\bibinfo{volume}{321}}, \bibinfo{pages}{1126} (\bibinfo{year}{2006}).

\bibitem[{\citenamefont{Gornyi et~al.}(2005)\citenamefont{Gornyi, Mirlin, and
  Polyakov}}]{GornyiPrl}
\bibinfo{author}{\bibfnamefont{I.~V.} \bibnamefont{Gornyi}},
  \bibinfo{author}{\bibfnamefont{A.~D.} \bibnamefont{Mirlin}},
  \bibnamefont{and} \bibinfo{author}{\bibfnamefont{D.~G.}
  \bibnamefont{Polyakov}}, \bibinfo{journal}{Phys. Rev. Lett.}
  \textbf{\bibinfo{volume}{95}}, \bibinfo{pages}{206603}
  (\bibinfo{year}{2005}).

\bibitem[{\citenamefont{Oganesyan and Huse}(2007)}]{OganesyanPrb}
\bibinfo{author}{\bibfnamefont{V.}~\bibnamefont{Oganesyan}} \bibnamefont{and}
  \bibinfo{author}{\bibfnamefont{D.~A.} \bibnamefont{Huse}},
  \bibinfo{journal}{Phys. Rev. B} \textbf{\bibinfo{volume}{75}},
  \bibinfo{pages}{155111} (\bibinfo{year}{2007}).

\bibitem[{\citenamefont{Iyer et~al.}(2013)\citenamefont{Iyer, Oganesyan,
  Refael, and Huse}}]{huseprb87}
\bibinfo{author}{\bibfnamefont{S.}~\bibnamefont{Iyer}},
  \bibinfo{author}{\bibfnamefont{V.}~\bibnamefont{Oganesyan}},
  \bibinfo{author}{\bibfnamefont{G.}~\bibnamefont{Refael}}, \bibnamefont{and}
  \bibinfo{author}{\bibfnamefont{D.~A.} \bibnamefont{Huse}},
  \bibinfo{journal}{Phys. Rev. B} \textbf{\bibinfo{volume}{87}},
  \bibinfo{pages}{134202} (\bibinfo{year}{2013}).

\bibitem[{\citenamefont{Basko et~al.}(2007)\citenamefont{Basko, Aleiner, and
  Altshuler}}]{baskoprb}
\bibinfo{author}{\bibfnamefont{D.~M.} \bibnamefont{Basko}},
  \bibinfo{author}{\bibfnamefont{I.~L.} \bibnamefont{Aleiner}},
  \bibnamefont{and} \bibinfo{author}{\bibfnamefont{B.~L.}
  \bibnamefont{Altshuler}}, \bibinfo{journal}{Phys. Rev. B}
  \textbf{\bibinfo{volume}{76}}, \bibinfo{pages}{052203}
  (\bibinfo{year}{2007}).

\bibitem[{\citenamefont{Ladieu et~al.}(1996)\citenamefont{Ladieu, Sanquer, and
  Bouchaud}}]{sanprb53}
\bibinfo{author}{\bibfnamefont{F.}~\bibnamefont{Ladieu}},
  \bibinfo{author}{\bibfnamefont{M.}~\bibnamefont{Sanquer}}, \bibnamefont{and}
  \bibinfo{author}{\bibfnamefont{J.~P.} \bibnamefont{Bouchaud}},
  \bibinfo{journal}{Phys. Rev. B} \textbf{\bibinfo{volume}{53}},
  \bibinfo{pages}{973} (\bibinfo{year}{1996}).

\bibitem[{\citenamefont{Goldman and Markovic}(1998)}]{goldmanpt51}
\bibinfo{author}{\bibfnamefont{A.~M.} \bibnamefont{Goldman}} \bibnamefont{and}
  \bibinfo{author}{\bibfnamefont{N.}~\bibnamefont{Markovic}},
  \bibinfo{journal}{Phys. Today} \textbf{\bibinfo{volume}{51}},
  \bibinfo{pages}{39} (\bibinfo{year}{1998}).

\bibitem[{\citenamefont{Gantmakher and Dolgopolov}(2010)}]{physupekhi}
\bibinfo{author}{\bibfnamefont{V.~F.} \bibnamefont{Gantmakher}}
  \bibnamefont{and} \bibinfo{author}{\bibfnamefont{V.~T.}
  \bibnamefont{Dolgopolov}}, \bibinfo{journal}{Phys.-Usp.}
  \textbf{\bibinfo{volume}{53}}, \bibinfo{pages}{1} (\bibinfo{year}{2010}).

\bibitem[{\citenamefont{Sondhi et~al.}(1997)\citenamefont{Sondhi, Girvin,
  Carini, and Shahar}}]{sondhirmp}
\bibinfo{author}{\bibfnamefont{S.~L.} \bibnamefont{Sondhi}},
  \bibinfo{author}{\bibfnamefont{S.~M.} \bibnamefont{Girvin}},
  \bibinfo{author}{\bibfnamefont{J.~P.} \bibnamefont{Carini}},
  \bibnamefont{and} \bibinfo{author}{\bibfnamefont{D.}~\bibnamefont{Shahar}},
  \bibinfo{journal}{Rev. Mod. Phys.} \textbf{\bibinfo{volume}{69}},
  \bibinfo{pages}{315} (\bibinfo{year}{1997}).

\bibitem[{\citenamefont{Hebard and Paalanen}(1990)}]{HebardPrl}
\bibinfo{author}{\bibfnamefont{A.~F.} \bibnamefont{Hebard}} \bibnamefont{and}
  \bibinfo{author}{\bibfnamefont{M.~A.} \bibnamefont{Paalanen}},
  \bibinfo{journal}{Phys. Rev. Lett.} \textbf{\bibinfo{volume}{65}},
  \bibinfo{pages}{927} (\bibinfo{year}{1990}).

\bibitem[{\citenamefont{Yazdani and Kapitulnik}(1995)}]{kapitulnikprl74}
\bibinfo{author}{\bibfnamefont{A.}~\bibnamefont{Yazdani}} \bibnamefont{and}
  \bibinfo{author}{\bibfnamefont{A.}~\bibnamefont{Kapitulnik}},
  \bibinfo{journal}{Phys. Rev. Lett.} \textbf{\bibinfo{volume}{74}},
  \bibinfo{pages}{3037} (\bibinfo{year}{1995}).

\bibitem[{\citenamefont{Baturina et~al.}(2004)\citenamefont{Baturina, Islamov,
  Bentner, Strunk, Baklanov, and Satta}}]{BaturinaJETP}
\bibinfo{author}{\bibfnamefont{T.~I.} \bibnamefont{Baturina}},
  \bibinfo{author}{\bibfnamefont{D.~R.} \bibnamefont{Islamov}},
  \bibinfo{author}{\bibfnamefont{J.}~\bibnamefont{Bentner}},
  \bibinfo{author}{\bibfnamefont{C.}~\bibnamefont{Strunk}},
  \bibinfo{author}{\bibfnamefont{M.~R.} \bibnamefont{Baklanov}},
  \bibnamefont{and} \bibinfo{author}{\bibfnamefont{A.}~\bibnamefont{Satta}},
  \bibinfo{journal}{JETP Lett.} \textbf{\bibinfo{volume}{79}},
  \bibinfo{pages}{337} (\bibinfo{year}{2004}).

\bibitem[{\citenamefont{Shahar and Ovadyahu}(1992)}]{Shaharprb}
\bibinfo{author}{\bibfnamefont{D.}~\bibnamefont{Shahar}} \bibnamefont{and}
  \bibinfo{author}{\bibfnamefont{Z.}~\bibnamefont{Ovadyahu}},
  \bibinfo{journal}{Phys. Rev. B} \textbf{\bibinfo{volume}{46}},
  \bibinfo{pages}{10917} (\bibinfo{year}{1992}).

\bibitem[{\citenamefont{Haviland et~al.}(1989)\citenamefont{Haviland, Liu, and
  Goldman}}]{haviprl62}
\bibinfo{author}{\bibfnamefont{D.~B.} \bibnamefont{Haviland}},
  \bibinfo{author}{\bibfnamefont{Y.}~\bibnamefont{Liu}}, \bibnamefont{and}
  \bibinfo{author}{\bibfnamefont{A.~M.} \bibnamefont{Goldman}},
  \bibinfo{journal}{Phys. Rev. Lett.} \textbf{\bibinfo{volume}{62}},
  \bibinfo{pages}{2180} (\bibinfo{year}{1989}).

\bibitem[{\citenamefont{Parendo et~al.}(2005)\citenamefont{Parendo, Tan,
  Bhattacharya, Eblen-Zayas, Staley, and Goldman}}]{goldmanprl94}
\bibinfo{author}{\bibfnamefont{K.~A.} \bibnamefont{Parendo}},
  \bibinfo{author}{\bibfnamefont{K.}~\bibnamefont{Tan}},
  \bibinfo{author}{\bibfnamefont{A.}~\bibnamefont{Bhattacharya}},
  \bibinfo{author}{\bibfnamefont{M.}~\bibnamefont{Eblen-Zayas}},
  \bibinfo{author}{\bibfnamefont{N.~E.} \bibnamefont{Staley}},
  \bibnamefont{and} \bibinfo{author}{\bibfnamefont{A.~M.}
  \bibnamefont{Goldman}}, \bibinfo{journal}{Phys. Rev. Lett.}
  \textbf{\bibinfo{volume}{94}}, \bibinfo{pages}{197004}
  (\bibinfo{year}{2005}).

\bibitem[{\citenamefont{Allain et~al.}(2012)\citenamefont{Allain, Han, and
  Bouchiat}}]{bounatmat11}
\bibinfo{author}{\bibfnamefont{A.}~\bibnamefont{Allain}},
  \bibinfo{author}{\bibfnamefont{Z.}~\bibnamefont{Han}}, \bibnamefont{and}
  \bibinfo{author}{\bibfnamefont{V.}~\bibnamefont{Bouchiat}},
  \bibinfo{journal}{Nat. Mater.} \textbf{\bibinfo{volume}{11}},
  \bibinfo{pages}{590} (\bibinfo{year}{2012}).

\bibitem[{\citenamefont{Sac\'ep\'e et~al.}(2008)\citenamefont{Sac\'ep\'e,
  Chapelier, Baturina, Vinokur, Baklanov, and Sanquer}}]{benjaminprl101}
\bibinfo{author}{\bibfnamefont{B.}~\bibnamefont{Sac\'ep\'e}},
  \bibinfo{author}{\bibfnamefont{C.}~\bibnamefont{Chapelier}},
  \bibinfo{author}{\bibfnamefont{T.~I.} \bibnamefont{Baturina}},
  \bibinfo{author}{\bibfnamefont{V.~M.} \bibnamefont{Vinokur}},
  \bibinfo{author}{\bibfnamefont{M.~R.} \bibnamefont{Baklanov}},
  \bibnamefont{and} \bibinfo{author}{\bibfnamefont{M.}~\bibnamefont{Sanquer}},
  \bibinfo{journal}{Phys. Rev. Lett.} \textbf{\bibinfo{volume}{101}},
  \bibinfo{pages}{157006} (\bibinfo{year}{2008}).

\bibitem[{\citenamefont{Crane et~al.}(2007)\citenamefont{Crane, Armitage,
  Johansson, Sambandamurthy, Shahar, and Gr\"uner}}]{craneprb752}
\bibinfo{author}{\bibfnamefont{R.~W.} \bibnamefont{Crane}},
  \bibinfo{author}{\bibfnamefont{N.~P.} \bibnamefont{Armitage}},
  \bibinfo{author}{\bibfnamefont{A.}~\bibnamefont{Johansson}},
  \bibinfo{author}{\bibfnamefont{G.}~\bibnamefont{Sambandamurthy}},
  \bibinfo{author}{\bibfnamefont{D.}~\bibnamefont{Shahar}}, \bibnamefont{and}
  \bibinfo{author}{\bibfnamefont{G.}~\bibnamefont{Gr\"uner}},
  \bibinfo{journal}{Phys. Rev. B} \textbf{\bibinfo{volume}{75}},
  \bibinfo{pages}{094506} (\bibinfo{year}{2007}).

\bibitem[{\citenamefont{Paalanen et~al.}(1992)\citenamefont{Paalanen, Hebard,
  and Ruel}}]{paalanenprl69}
\bibinfo{author}{\bibfnamefont{M.~A.} \bibnamefont{Paalanen}},
  \bibinfo{author}{\bibfnamefont{A.~F.} \bibnamefont{Hebard}},
  \bibnamefont{and} \bibinfo{author}{\bibfnamefont{R.~R.} \bibnamefont{Ruel}},
  \bibinfo{journal}{Phys. Rev. Lett.} \textbf{\bibinfo{volume}{69}},
  \bibinfo{pages}{1604} (\bibinfo{year}{1992}).

\bibitem[{\citenamefont{Gantmakher et~al.}(1996)\citenamefont{Gantmakher,
  Golubkov, Lok, and Geim}}]{GantmakherJETP}
\bibinfo{author}{\bibfnamefont{V.~F.} \bibnamefont{Gantmakher}},
  \bibinfo{author}{\bibfnamefont{M.~V.} \bibnamefont{Golubkov}},
  \bibinfo{author}{\bibfnamefont{J.~G.~S.} \bibnamefont{Lok}},
  \bibnamefont{and} \bibinfo{author}{\bibfnamefont{A.~K.} \bibnamefont{Geim}},
  \bibinfo{journal}{JETP} \textbf{\bibinfo{volume}{82}}, \bibinfo{pages}{951}
  (\bibinfo{year}{1996}).

\bibitem[{\citenamefont{Sambandamurthy
  et~al.}(2004)\citenamefont{Sambandamurthy, Engel, Johansson, and
  Shahar}}]{murthyprl2004}
\bibinfo{author}{\bibfnamefont{G.}~\bibnamefont{Sambandamurthy}},
  \bibinfo{author}{\bibfnamefont{L.~W.} \bibnamefont{Engel}},
  \bibinfo{author}{\bibfnamefont{A.}~\bibnamefont{Johansson}},
  \bibnamefont{and} \bibinfo{author}{\bibfnamefont{D.}~\bibnamefont{Shahar}},
  \bibinfo{journal}{Phys. Rev. Lett.} \textbf{\bibinfo{volume}{92}},
  \bibinfo{pages}{107005} (\bibinfo{year}{2004}).

\bibitem[{\citenamefont{Nguyen et~al.}(2009)\citenamefont{Nguyen, Hollen,
  Stewart, Shainline, Yin, Xu, and Valles}}]{vallesprl103}
\bibinfo{author}{\bibfnamefont{H.~Q.} \bibnamefont{Nguyen}},
  \bibinfo{author}{\bibfnamefont{S.~M.} \bibnamefont{Hollen}},
  \bibinfo{author}{\bibfnamefont{M.~D.} \bibnamefont{Stewart}},
  \bibinfo{author}{\bibfnamefont{J.}~\bibnamefont{Shainline}},
  \bibinfo{author}{\bibfnamefont{A.}~\bibnamefont{Yin}},
  \bibinfo{author}{\bibfnamefont{J.~M.} \bibnamefont{Xu}}, \bibnamefont{and}
  \bibinfo{author}{\bibfnamefont{J.~M.} \bibnamefont{Valles}},
  \bibinfo{journal}{Phys. Rev. Lett.} \textbf{\bibinfo{volume}{103}},
  \bibinfo{pages}{157001} (\bibinfo{year}{2009}).

\bibitem[{\citenamefont{Feigel'man
  et~al.}(2010{\natexlab{a}})\citenamefont{Feigel'man, Ioffe, Kravtsov, and
  Cuevas}}]{FeigAnnals}
\bibinfo{author}{\bibfnamefont{M.}~\bibnamefont{Feigel'man}},
  \bibinfo{author}{\bibfnamefont{L.}~\bibnamefont{Ioffe}},
  \bibinfo{author}{\bibfnamefont{V.}~\bibnamefont{Kravtsov}}, \bibnamefont{and}
  \bibinfo{author}{\bibfnamefont{E.}~\bibnamefont{Cuevas}},
  \bibinfo{journal}{Annals of Phys.} \textbf{\bibinfo{volume}{325}},
  \bibinfo{pages}{1390} (\bibinfo{year}{2010}{\natexlab{a}}).

\bibitem[{\citenamefont{Dubi et~al.}(2007)\citenamefont{Dubi, Meir, and
  Avishai}}]{YonatanNat}
\bibinfo{author}{\bibfnamefont{Y.}~\bibnamefont{Dubi}},
  \bibinfo{author}{\bibfnamefont{Y.}~\bibnamefont{Meir}}, \bibnamefont{and}
  \bibinfo{author}{\bibfnamefont{Y.}~\bibnamefont{Avishai}},
  \bibinfo{journal}{Nature} \textbf{\bibinfo{volume}{449}},
  \bibinfo{pages}{876} (\bibinfo{year}{2007}).

\bibitem[{\citenamefont{Sambandamurthy
  et~al.}(2005)\citenamefont{Sambandamurthy, Engel, Johansson, Peled, and
  Shahar}}]{murthyprl}
\bibinfo{author}{\bibfnamefont{G.}~\bibnamefont{Sambandamurthy}},
  \bibinfo{author}{\bibfnamefont{L.~W.} \bibnamefont{Engel}},
  \bibinfo{author}{\bibfnamefont{A.}~\bibnamefont{Johansson}},
  \bibinfo{author}{\bibfnamefont{E.}~\bibnamefont{Peled}}, \bibnamefont{and}
  \bibinfo{author}{\bibfnamefont{D.}~\bibnamefont{Shahar}},
  \bibinfo{journal}{Phys. Rev. Lett.} \textbf{\bibinfo{volume}{94}},
  \bibinfo{pages}{017003} (\bibinfo{year}{2005}).

\bibitem[{\citenamefont{Baturina et~al.}(2007)\citenamefont{Baturina, Mironov,
  Vinokur, Baklanov, and Strunk}}]{BaturinaPRL}
\bibinfo{author}{\bibfnamefont{T.~I.} \bibnamefont{Baturina}},
  \bibinfo{author}{\bibfnamefont{A.~Y.} \bibnamefont{Mironov}},
  \bibinfo{author}{\bibfnamefont{V.~M.} \bibnamefont{Vinokur}},
  \bibinfo{author}{\bibfnamefont{M.~R.} \bibnamefont{Baklanov}},
  \bibnamefont{and} \bibinfo{author}{\bibfnamefont{C.}~\bibnamefont{Strunk}},
  \bibinfo{journal}{Phys. Rev. Lett.} \textbf{\bibinfo{volume}{99}},
  \bibinfo{pages}{257003} (\bibinfo{year}{2007}).

\bibitem[{\citenamefont{Kalok et~al.}(2012)\citenamefont{Kalok, Bilusic,
  Baturina, Mironov, Postolova, amd A.~V.~Latyshev, Vinokur, and
  Strunk}}]{KalokJP}
\bibinfo{author}{\bibfnamefont{D.}~\bibnamefont{Kalok}},
  \bibinfo{author}{\bibfnamefont{A.}~\bibnamefont{Bilusic}},
  \bibinfo{author}{\bibfnamefont{T.~I.} \bibnamefont{Baturina}},
  \bibinfo{author}{\bibfnamefont{A.~Y.} \bibnamefont{Mironov}},
  \bibinfo{author}{\bibfnamefont{S.~V.} \bibnamefont{Postolova}},
  \bibinfo{author}{\bibfnamefont{A.~K.~G.} \bibnamefont{amd A.~V.~Latyshev}},
  \bibinfo{author}{\bibfnamefont{V.~M.} \bibnamefont{Vinokur}},
  \bibnamefont{and} \bibinfo{author}{\bibfnamefont{C.}~\bibnamefont{Strunk}},
  \bibinfo{journal}{J. Phys.:Conference Series} \textbf{\bibinfo{volume}{400}},
  \bibinfo{pages}{022042} (\bibinfo{year}{2012}).

\bibitem[{\citenamefont{Vinokur et~al.}(2008)\citenamefont{Vinokur, Baturina,
  Fistul, Mironov, Baklanov, and Strunk}}]{vinokurnat}
\bibinfo{author}{\bibfnamefont{V.~M.} \bibnamefont{Vinokur}},
  \bibinfo{author}{\bibfnamefont{T.~I.} \bibnamefont{Baturina}},
  \bibinfo{author}{\bibfnamefont{M.~V.} \bibnamefont{Fistul}},
  \bibinfo{author}{\bibfnamefont{A.~Y.} \bibnamefont{Mironov}},
  \bibinfo{author}{\bibfnamefont{M.~R.} \bibnamefont{Baklanov}},
  \bibnamefont{and} \bibinfo{author}{\bibfnamefont{C.}~\bibnamefont{Strunk}},
  \bibinfo{journal}{Nature} \textbf{\bibinfo{volume}{452}},
  \bibinfo{pages}{613} (\bibinfo{year}{2008}).

\bibitem[{\citenamefont{Altshuler et~al.}(2009)\citenamefont{Altshuler,
  Kravtsov, Lerner, and Aleiner}}]{borisprl}
\bibinfo{author}{\bibfnamefont{B.~L.} \bibnamefont{Altshuler}},
  \bibinfo{author}{\bibfnamefont{V.~E.} \bibnamefont{Kravtsov}},
  \bibinfo{author}{\bibfnamefont{I.~V.} \bibnamefont{Lerner}},
  \bibnamefont{and} \bibinfo{author}{\bibfnamefont{I.~L.}
  \bibnamefont{Aleiner}}, \bibinfo{journal}{Phys. Rev. Lett.}
  \textbf{\bibinfo{volume}{102}}, \bibinfo{pages}{176803}
  (\bibinfo{year}{2009}).

\bibitem[{\citenamefont{Ovadia et~al.}(2009)\citenamefont{Ovadia, Sacepe, and
  Shahar}}]{maozprl}
\bibinfo{author}{\bibfnamefont{M.}~\bibnamefont{Ovadia}},
  \bibinfo{author}{\bibfnamefont{B.}~\bibnamefont{Sacepe}}, \bibnamefont{and}
  \bibinfo{author}{\bibfnamefont{D.}~\bibnamefont{Shahar}},
  \bibinfo{journal}{Phys. Rev. Lett.} \textbf{\bibinfo{volume}{102}},
  \bibinfo{pages}{176802} (\bibinfo{year}{2009}).

\bibitem[{\citenamefont{Shtyk et~al.}(2013)\citenamefont{Shtyk, Feigel'man, and
  Kravtsov}}]{kravtsovprl111}
\bibinfo{author}{\bibfnamefont{A.~V.} \bibnamefont{Shtyk}},
  \bibinfo{author}{\bibfnamefont{M.~V.} \bibnamefont{Feigel'man}},
  \bibnamefont{and} \bibinfo{author}{\bibfnamefont{V.~E.}
  \bibnamefont{Kravtsov}}, \bibinfo{journal}{Phys. Rev. Lett.}
  \textbf{\bibinfo{volume}{111}}, \bibinfo{pages}{166603}
  (\bibinfo{year}{2013}).

\bibitem[{\citenamefont{Efros and Shklovskii}(1975)}]{efros}
\bibinfo{author}{\bibfnamefont{A.~L.} \bibnamefont{Efros}} \bibnamefont{and}
  \bibinfo{author}{\bibfnamefont{B.~I.} \bibnamefont{Shklovskii}},
  \bibinfo{journal}{J. Phys. C} \textbf{\bibinfo{volume}{8}},
  \bibinfo{pages}{L49} (\bibinfo{year}{1975}).

\bibitem[{\citenamefont{Baturina et~al.}(2008)\citenamefont{Baturina, Mironov,
  Vinokur, Baklanov, and Strunk}}]{baturinajetp88}
\bibinfo{author}{\bibfnamefont{T.~I.} \bibnamefont{Baturina}},
  \bibinfo{author}{\bibfnamefont{A.~Y.} \bibnamefont{Mironov}},
  \bibinfo{author}{\bibfnamefont{V.~M.} \bibnamefont{Vinokur}},
  \bibinfo{author}{\bibfnamefont{M.~R.} \bibnamefont{Baklanov}},
  \bibnamefont{and} \bibinfo{author}{\bibfnamefont{C.}~\bibnamefont{Strunk}},
  \bibinfo{journal}{JETP Lett.} \textbf{\bibinfo{volume}{88}},
  \bibinfo{pages}{752} (\bibinfo{year}{2008}).

\bibitem[{\citenamefont{Gopalakrishnan and Nandkishore}(2014)}]{RahulArx}
\bibinfo{author}{\bibfnamefont{S.}~\bibnamefont{Gopalakrishnan}}
  \bibnamefont{and}
  \bibinfo{author}{\bibfnamefont{R.}~\bibnamefont{Nandkishore}},
  \bibinfo{journal}{Phys. Rev. B} \textbf{\bibinfo{volume}{90}},
  \bibinfo{pages}{224203} (\bibinfo{year}{2014}).

\bibitem[{\citenamefont{Ben-Chorin et~al.}(1993)\citenamefont{Ben-Chorin,
  Ovadyahu, and Pollak}}]{ov1}
\bibinfo{author}{\bibfnamefont{M.}~\bibnamefont{Ben-Chorin}},
  \bibinfo{author}{\bibfnamefont{Z.}~\bibnamefont{Ovadyahu}}, \bibnamefont{and}
  \bibinfo{author}{\bibfnamefont{M.}~\bibnamefont{Pollak}},
  \bibinfo{journal}{Phys. Rev. B} \textbf{\bibinfo{volume}{48}},
  \bibinfo{pages}{15025} (\bibinfo{year}{1993}).

\bibitem[{\citenamefont{Ovadyahu}(2012)}]{ov2}
\bibinfo{author}{\bibfnamefont{Z.}~\bibnamefont{Ovadyahu}},
  \bibinfo{journal}{Phys. Rev. Lett.} \textbf{\bibinfo{volume}{108}},
  \bibinfo{pages}{156602} (\bibinfo{year}{2012}).

\bibitem[{\citenamefont{Ovadia et~al.}(2013)\citenamefont{Ovadia, Kalok,
  Sacepe, and Shahar}}]{MaozNat}
\bibinfo{author}{\bibfnamefont{M.}~\bibnamefont{Ovadia}},
  \bibinfo{author}{\bibfnamefont{D.}~\bibnamefont{Kalok}},
  \bibinfo{author}{\bibfnamefont{B.}~\bibnamefont{Sacepe}}, \bibnamefont{and}
  \bibinfo{author}{\bibfnamefont{D.}~\bibnamefont{Shahar}},
  \bibinfo{journal}{Nat. Phys} \textbf{\bibinfo{volume}{9}},
  \bibinfo{pages}{415} (\bibinfo{year}{2013}).

\bibitem[{\citenamefont{Feigel'man
  et~al.}(2010{\natexlab{b}})\citenamefont{Feigel'man, Ioffe, and
  M\'ezard}}]{FeigPrb}
\bibinfo{author}{\bibfnamefont{M.~V.} \bibnamefont{Feigel'man}},
  \bibinfo{author}{\bibfnamefont{L.~B.} \bibnamefont{Ioffe}}, \bibnamefont{and}
  \bibinfo{author}{\bibfnamefont{M.}~\bibnamefont{M\'ezard}},
  \bibinfo{journal}{Phys. Rev. B} \textbf{\bibinfo{volume}{82}},
  \bibinfo{pages}{184534} (\bibinfo{year}{2010}{\natexlab{b}}).

\end{thebibliography}

\end{document}